# Solar Cycle Variation of Sustained Gamma-ray Emission Events from the Sun and Related Energetic Events

*Gopalswamy N.[1], Mäkelä P.A.[1,2], Akiyama S.[1,2], Yashiro S.[1,2], Xie H.[1,2]*

[1]NASA Goddard Space Flight Center, Greenbelt, MD 20771, USA; nat.gopalswamy@nasa.gov
[2]The Catholic University of America, Washington DC 20064

1. **Abstract**.

The sustained gamma-ray emission (SGRE) from the Sun is one of the fascinating high-energy phenomena closely related to the acceleration of protons to energies >300 MeV. Here we report on the solar cycle variation of SGRE events based on observations from Fermi's Large Area Telescope (LAT). This report covers solar cycles (SCs) 24 and 25 during which Fermi has been operating. Since SGRE events are closely related to solar energetic particle (SEP) events and interplanetary type II radio bursts caused by fast and wide coronal mass ejections (CMEs), we consider these phenomena as well. Many studies have shown that SC 25 is similar or slightly stronger than SC 24. The number of SEP events, GLE events, IP type II bursts, and fast and wide CMEs confirm this conclusion. However, the number of SGRE events observed by Fermi/LAT has diminished significantly in SC 25 relative to SC 24. One of the issues has been the reduced coverage of the Sun since 2018 due to a mechanical problem with a solar array of the Fermi mission. By identifying the Fermi/LAT gaps and the number of energetic events (fast and wide CMEs, interplanetary type II bursts) we conclude that about three times more SGRE events must have occurred than the 15 events observed by Fermi.

2. **Introduction**

The sustained gamma-ray emission (SGRE) from the Sun lasts well beyond the impulsive phase of the associated solar flare. The emission is due to the decay of neutral pions suggesting that the underlying >300 MeV particles might have the same origin as large solar energetic particle (SEP) events [Forrest et al. 1985]. For more than two decades since their discovery, only a handful SGRE events were detected by various instruments [see Chupp and Ryan 2009]. The situation changed when the Fermi satellite was launched in 2008. Fermi's Large Area Telescope (LAT, Atwood et al. [2009]) detected dozens of events in solar cycles (SCs) 24 and 25 [Share et al. 2018; Allafort 2018; Ajello et al. 2021]. These observations show a close relation among SGRE events, interplanetary (IP) type II bursts, SEP events, and energetic (fast and wide) coronal mass ejections (CMEs). The physical connection can be understood as follows: energetic CMEs drive strong shocks that accelerate electrons and protons. The accelerated electrons give rise to type II radio bursts, while the >300 MeV protons propagate to the Sun from the shock downstream, interact with the dense atmosphere, and produce neutral pions that decay into gamma-rays. Energetic particles from the shock upstream escape into space to be detected as SEP events by spacecraft. CMEs underlying SGRE events possess properties similar to CMEs producing ground level enhancement (GLE) in SEP events [Gopalswamy et al. 2018]. Many shocks continue to accelerate particles until Earth and even beyond and thus are capable of supplying energetic particles over a long period of time to sustain the gamma-ray emission. This is also indicated by the ending frequency of the associated IP type II bursts indicating shocks efficiently accelerating electrons survive for long distances from the Sun. One of the implications for the shock source of >300 MeV protons is that the SGRE source needs to be extended – tens of degrees – around the eruption site. The occurrence of SGRE in association with eruptions occurring tens of degrees behind the limb strongly support the shock paradigm [Pesce-Rollins et al. 2015; Plotnikov et al. 2017; Jin et al. 2017; Gopalswamy et al. 2020; 2025]. Another potential source of >300 MeV protons is the flare reconnection region. Particles accelerated in the reconnection region can get trapped into post-eruption arcade (PEA) loops and can precipitate to the photosphere causing impulsive gamma-rays [Ryan and Lee 1991; de Nolfo et al. 2019]. These sources are necessarily compact (a maximum extent of ~15 degrees). Sustaining energetic protons in such loops to account of the long-duration SGRE requires unusual turbulence conditions [Kanbach et al. 1993]. In this paper, we take CME-driven shocks as the source of energetic particles responsible for SGRE.

In section 3, we compare the SGRE and associated events occurring during the first 61 months of SCs 24 (December 2008 to December 2013) and 25 (December 2019 to December 2024). These epochs roughly include the rise and maximum phases of the two cycles. In section 4, we estimate the number of SGRE events, considering the sunspot number, SEP events, fast and wide (FW) CMEs, and IP type II bursts identified during Fermi/LAT data gaps that started in March 2018. In section 5, we provide our summary and conclusions.

### 3. Observations

Fermi/LAT observations are available at https://hesperia.gsfc.nasa.gov/fermi_solar/ up to date and in the published catalogs for SC 24 [Share et al. 2018, Allafort 2018, and Ajello et al. 2021]. CME, SEP, and type II burst data are from the CDAW data center [https://cdaw.gsfc.nasa.gov, Yashiro et al. 2004; Gopalswamy et al. 2009; 2010; 2019; Gopalswamy et al. 2024]. Details on the solar array drive anomaly (SODA) including various observing modes employed are available in the Fermi web site (https://fermi.gsfc.nasa.gov/ssc/observations/types/post_anomaly/). In order to extract the times of Sun exposure, we make use of the information provided in the Fermi Timeline web site: https://fermi.gsfc.nasa.gov/ssc/observations/timeline/posting/. For solar cycle reference, we use the sunspot number (SSN) V2.0 available from the Sunspot Index and Long-term Solar Observations (SILSO) maintained at https://wwwbis.sidc.be/silso/infosnmtot. Solar wind data to characterize the solar cycle variation of the heliospheric state are obtained from NASA's OMNI database (https://omniweb.gsfc.nasa.gov/).

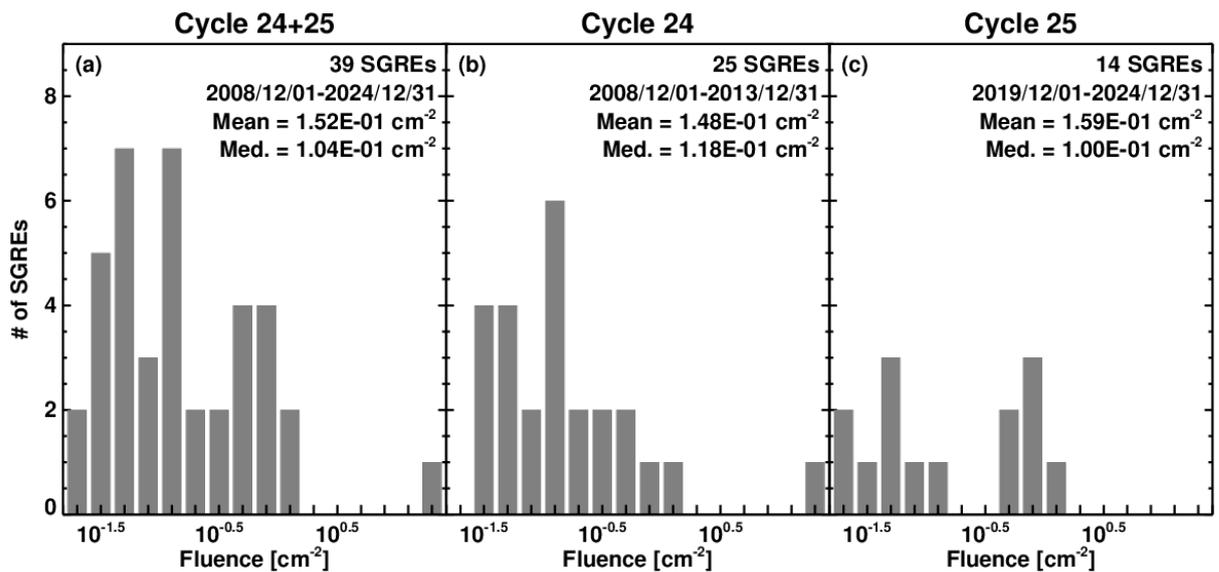

**Figure 1.** *Histograms of SGRE fluence in SCs 24 and 25 (a), SC 24 (b), and SC 25 (c).*

Figure 1 shows fluence distributions of SGRE events that occurred during the first 61 months in SCs 24 and 25. The number of SGRE events in SC 24 is much higher in SC 25. The mean values are very similar in the two cycles, but smaller than the mean value (0.49 cm$^{-2}$) over the whole of SC 24 [Gopalswamy et al. 2019]. Thus, the primary difference seem to be the number of events between the two cycles. Figure 2 shows the histograms of the SGRE durations over the first 61 months in SCs 24 and 25. The duration distributions are similar, although the mean values are slightly smaller in SC 25. Over the whole of cycle 24, the average duration is larger by a factor of 2. The duration comparison is probably influenced by the large SODA-related data gaps in Fermi/LAT observations since March 2018 (see later).

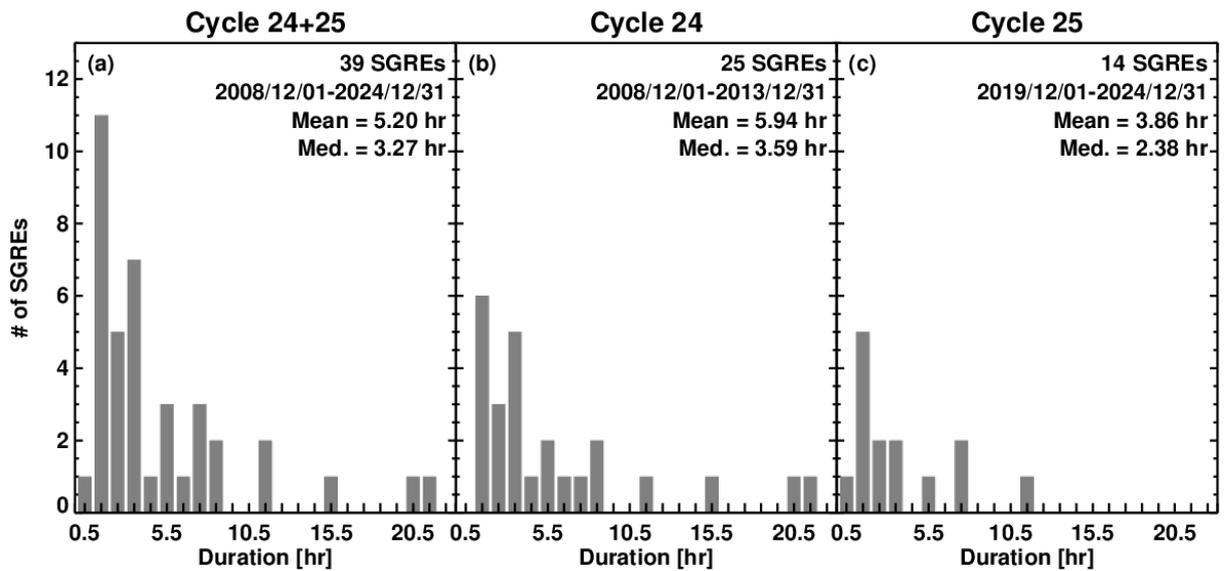

**Figure 2**. *Histograms of SGRE durations in SCs 24 and 25 (a), SC 24 (b), and SC 25 (c).*

### 3.1 Overview of SGRE Events

Table 1 compares the number of energetic events related to SGRE between SCs 24 and 25 over the first 61 months in each cycle: CMEs, SEPs, geomagnetic storms, decameter-hectometric (DH) type II bursts, and solar wind properties. The solar wind properties are listed because they can influence the properties of CMEs and their interaction with the solar wind medium and hence their ability in causing SEPs and geomagnetic storms. The SILSO sunspot number (SSN) averaged over the first 61 months are 56.9 and 79.0, respectively, indicating that SC 25 is ~39% stronger than SC 24, opposite of what happened between SCs 23 and 24 [Gopalswamy et al. 2014; 2015a]. The number of SGRE events shows a decline of ~44% in SC 25 with a similar decline (~40%) when normalized to SSN. SC 25 has been very active at the maximum phase, with a significantly larger number of halo CMEs observed in SC 25. However, the halo CME abundance normalized to the SSN decreased by ~7%, which is consistent with the fact that weaker cycles have a higher halo CME abundance [Gopalswamy et al. 2015b]. The number of FW (speed V≥900 km/s and width W≥60º) CMEs and DH type II bursts normalized to the SSN increased by 29% and 33%, respectively in SC 25. The number of major flares (soft X-ray flare class M1.0 or higher) normalized to the SSN increased by 182%. The number of large SEP events (proton intensity ≥10 pfu) in SC 25 decreased by 17% when normalized to SSN. However, the number of GLE events in SC 25 quadrupled compared to SC 24, indicating that normalized number shows an increase of 150%. The number of intense geomagnetic storms (Dst ≤ -100 nT) in SC 25 increased by 50%, although when normalized to SSN, the increase is

~ 6%. Halo CMEs, FW CMEs, SEPs, GLEs, and DH type II bursts are also closely associated with SGREs. Of these halo CMEs are influenced by heliospheric pressure, while SEPs are affected by the magnetic connectivity of the eruption region to GOES. The remaining ones, viz., FW CMEs and DH type II bursts have similar solar cycle variation because DH type II bursts are produced by FW CMEs [Gopalswamy et al. 2001]. While all the energetic events indicate a stronger cycle 25, the SGRE events has the opposite trend. We expect the number of SGREs roughly follow the solar cycle variation of FW CMEs and DH type II bursts. The number of SGRE events normalized to the number of DH type II bursts is 0.25 in SC 24, whereas it drops to 0.08 in SC 25. This is clearly abnormal, which we analyze in detail in the following subsections.

*Table 1. Number (#) of energetic event and solar wind properties in Solar Cycles 24 and 25*

| Property | SC 24 | SC 25 | Ratio |
|---|---|---|---|
| Averaged SSN | 56.9 | 79.0 | 1.39 |
| #SGRE events | 25 (0.44) | 14 (0.18) | 0.56 (0.4) |
| #Halo CMEs | 192 (3.37)[a] | 247 (3.12) | 1.29 (0.93) |
| # FW CMEs V≥900 km/s & W≥60° | 149 (2.62) | 268 (3.39) | 1.80 (1.29) |
| #DH Type II bursts | 99 (1.74) | 183 (2.32) | 1.85 (1.33) |
| #≥M1.0 flares | 389 (6.84) | 1525 (19.30) | 3.92 (2.82) |
| #≥10 pfu SEP events | 30 (0.53) | 35 (0.44) | 1.17 (0.83) |
| # GLE events | 1 (0.02) | 4 (0.05) | 4.0 (2.50) |
| # Intense storms (Dst ≤ -100 nT) | 12 (0.21) | 18 (0.23) | 1.5 (1.06) |
| #SGRE/#FW CMEs | 0.17 | 0.05 | 0.31 |
| #SGRE/#DH type II | 0.25 | 0.08 | 0.30 |
| #SGRE/#SEP events | 0.83 | 0.40 | 0.48 |
| Total Pressure (pPa) | 27.1 | 32.6 | 1.20 |
| Magnetic field strength B (nT) | 4.92 | 5.50 | 1.12 |
| Bulk speed V km/s | 399 | 410 | 1.03 |
| Proton density N (cm$^{-3}$) | 5.07 | 5.62 | 1.11 |
| Proton temperature T ($10^5$ K) | 0.75 | 0.88 | 1.18 |
| Alfven Speed $V_A$ (km/s) | 47.9 | 50.9 | 1.06 |

[a]The numbers within parentheses are normalized to the average SSN

### 3.2 Solar Wind Parameters

The number of energetic events such as GLEs and intense geomagnetic storms is known to decrease more drastically than SSN when the heliospheric state is weak [Gopalswamy et al. 2014; 2015a]. The heliospheric state is discerned from in-situ observations made by spacecraft at Sun-Earth L1. The dominant factor that characterizes the heliospheric state is the total pressure (magnetic + gas) in the heliosphere. We made use of OMNI data to derive the total pressure from the observations of magnetic field strength, solar wind plasma density, and temperature. The monthly averages of the derived/measured quantities are plotted in Figure 3: solar wind bulk speed, total pressure, magnetic field, density, temperature, and Alfven speed. We see that the total pressure averaged over the first 61 months in SC 25 increased by 20% compared to SC 24, somewhat smaller than the increase in SSN. The combined increases of solar wind magnetic field, density and temperature can readily account for the increase in total pressure. With higher heliospheric pressure, one does not expect energetic events to decline in SC 25. In fact, we see an increase in the number of energetic events such as GLEs and intense geomagnetic storms (Table 1). The reduction in the halo CME abundance normalized to SSN in SC 25 is consistent with the increased heliospheric pressure [Gopalswamy et al. 2014; 2015b]. While the solar wind speed did not change significantly, the Alfven speed increased by ~6%. If the higher Alfven speed measured at 1 au applies to the near-Sun ambient medium, one expects that it is harder to form shocks and hence might have contributed partially to the reduction in the normalized number of large SEP events in SC 25.

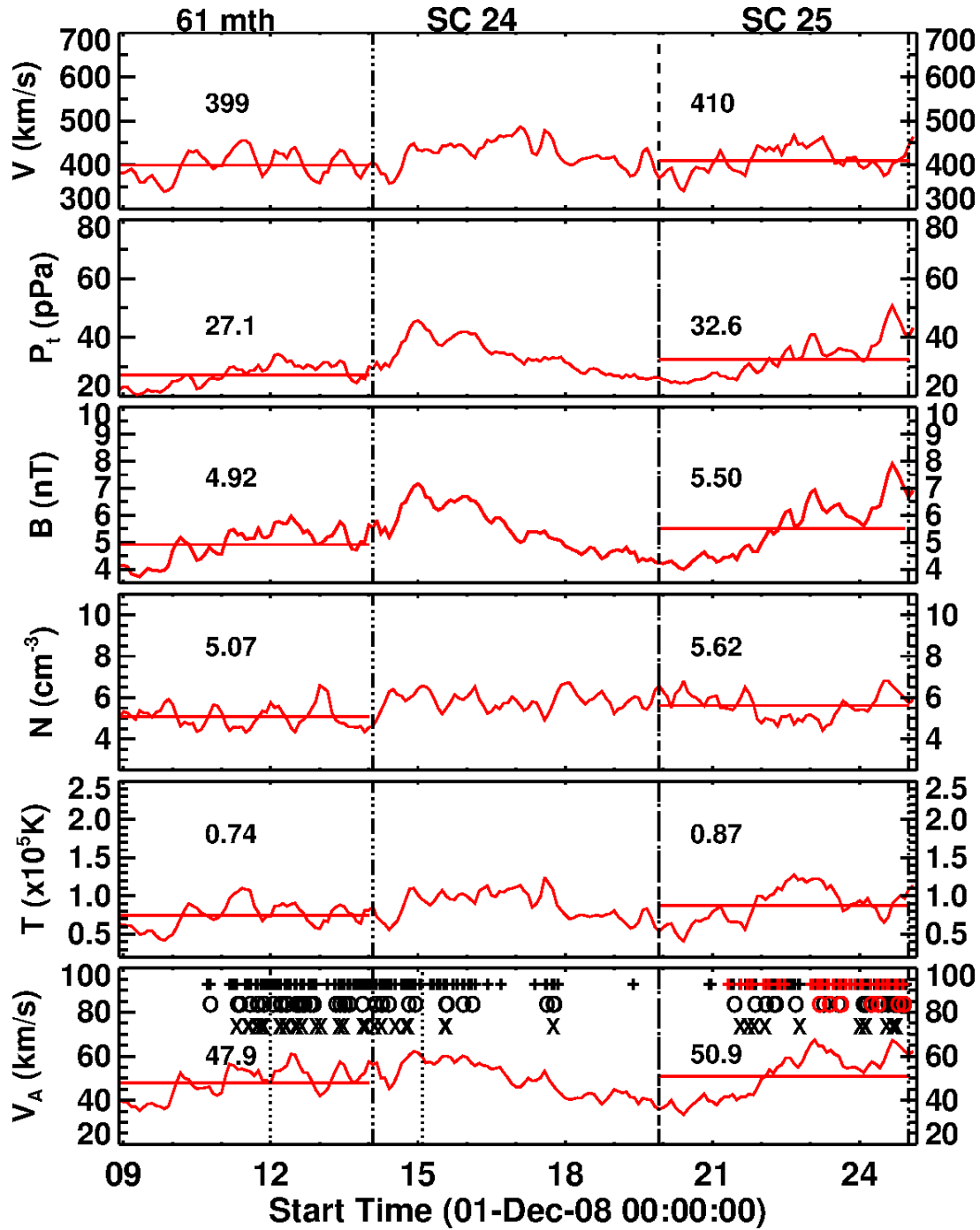

*Figure 3*. Solar wind parameters from 2009 to 2024: bulk speed (V), total pressure (Pt), magnetic field strength (B), density (N), temperature (T), Alfven speed ($V_A$). The corresponding epochs (first 61 months) in the two cycles are indicated by the horizontal red lines and the average value of each parameter over the epochs is noted and listed in Table 1. For example, the average $V_A$ is 48.2 km/s in SC 24 and 51.1 km/s in SC 25. The three rows of symbols in the bottom ($V_A$) panel mark occurrence dates of DH type II bursts (+), SEP events (O) and SGRE events (X). The symbols in red denote events occurring during Fermi/LAT data gaps. The vertical dashed lines in the bottom panel mark the maximum phase in SC 24.

Topic:

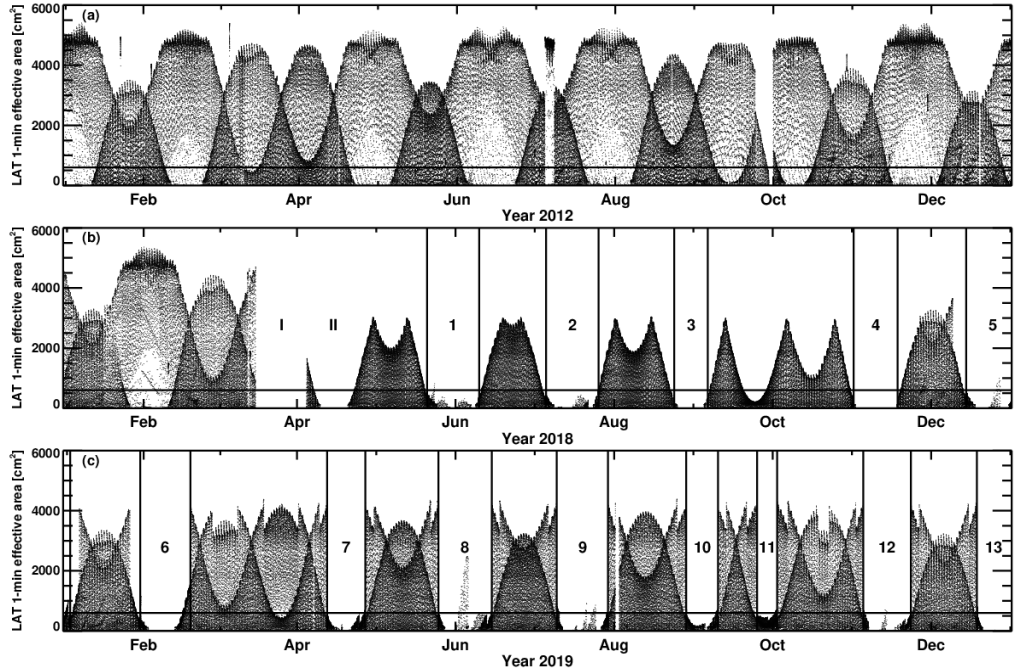

*Figure 4.* Plots of solar exposures during the years 2012 (a), 2018 (b), and 2019 (c). The horizontal lines in each panel at a 1-min exposure area of 650 $cm^2$ represents the cutoff value. Note that the exposure gaps start in 2018 March. Gaps (I) and (II) in March and April 2018 represent the transition period before new observing modes are established. Gaps 1-13 represent the lack of solar exposure in the remaining months of 2018 and the whole of 2019. The complicated exposure profile is due to Fermi using different survey modes (two-sided rocking (+/- 50 deg survey or 50/60 deg asymmetric survey) against the rate for modified sine observations) interleaved throughout the year depending on the position of the Sun relative to the orbit plane. During the exposure gaps, the modified sine is used.

### *3.3 Fermi/LAT Data Coverage in SC 25*

In this section we examine Fermi/LAT timeline of observations to investigate the SODA-related reduction in solar coverage that started in 2018. Although the solar array is fully functional, the solar array in question got stuck at an angle to the LAT boresight because the drive assembly is not able to rotate the array. In order for the stuck array to be illuminated, the Sun is kept toward the edge of the LAT field of view resulting in a reduced exposure with LAT. After identifying the intervals of SODA-related data gaps, we also identify energetic events such as FW CMEs, DH type II bursts, and large SEP events.

Figure 4 illustrates the reduction in solar exposure using plots for the pre-anomaly year 2012, the anomaly year 2018, and the year 2019 that is fully under the reduced capability. Comparing the plots for the year 2012 (panel a) and 2019 (panel c) in Fig. 4, we see two important changes. (i) there are SODA-related large gaps (1-3 weeks at a time) in the Sun exposure, and (ii) the maximum LAT effective area is reduced to below 4000 $cm^2$ in 2019 and later years compared to a maximum of 5500 $cm^2$ in the pre-anomaly period. The gaps are substantial, so any SGRE event occurring during these gaps will not be counted. We also manually examined the Fermi/LAT 4-day quick look plots (https://hesperia.gsfc.nasa.gov/fermi_solar/) to identify extended intervals with no Sun exposure. We see that there are 8 intervals of large data gap in the year 2019. These intervals are consistent with the modified sine observations during which there was no Sun exposure. For example, the gap marked 6 in panel Fig. 4c started at 19:44 UT on 23 January 2019 and ended at 03:52 UT on 16 February 2019, amounting a gap of ~24 days.

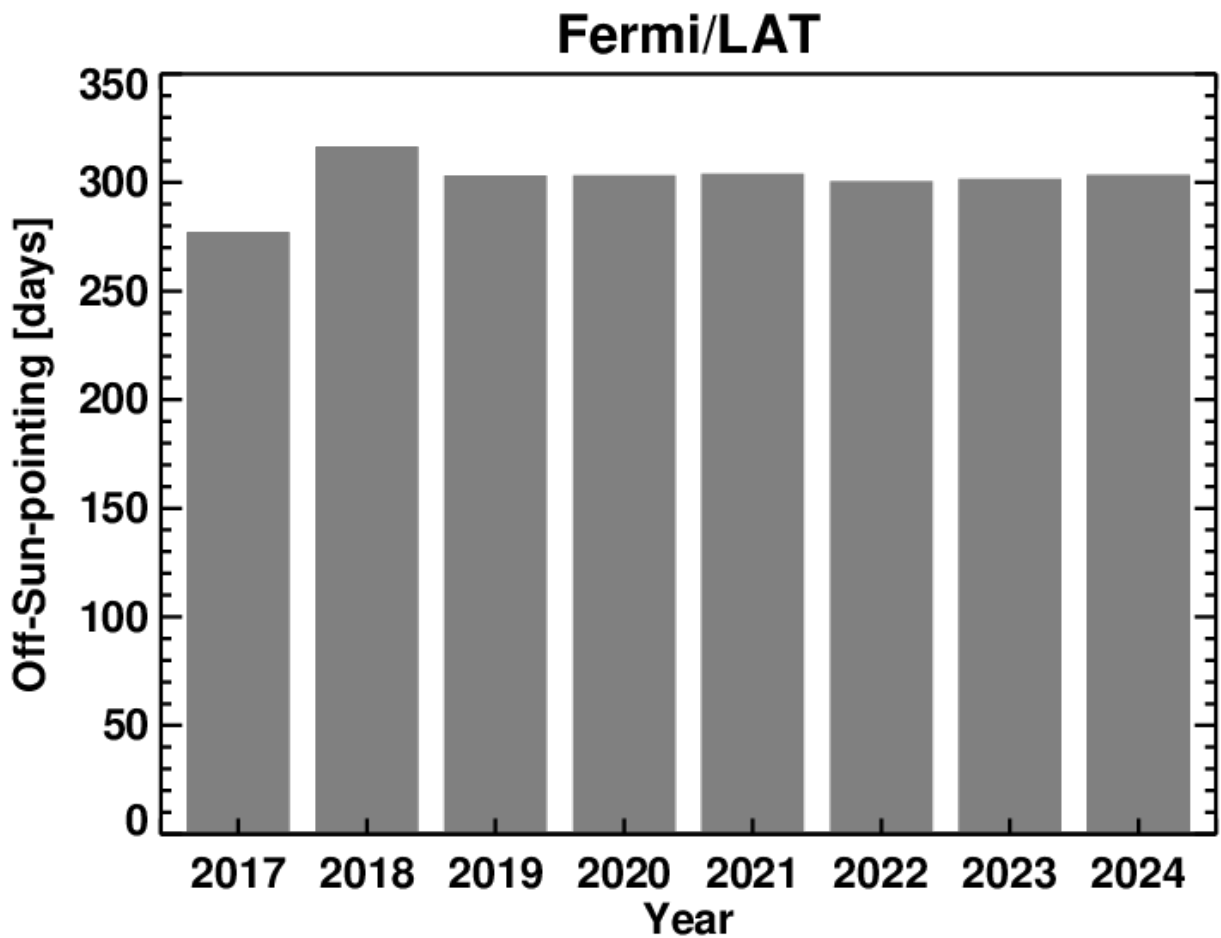

***Figure 5.*** *Effective number of days per year during which the Sun had no Fermi/LAT exposure. As before, the data for the pre-anomaly year 2017 is compared with the post-anomaly years. In 2017, there was no Sun exposure for ~275 days. In 2018, new observing mode was being established, so there was no Sun exposure for ~315 days. During later years (2019-2024), there were no Sun exposures for ~305 days. Note that the normal level of Sun exposure during the pre-anomaly years is ~90 days per year because of the 96 min orbit (only 48 min on the dayside) during which the Sun exposure is typically for ~30 min. This is the reason for Fermi/LAT having no Sun exposure for 275 days in the pre-anomaly years. In other words, the effective number of days that Fermi/LAT had Sun exposure decreased from ~90 days per year to ~60 days per year because of the SODA incident.*

Counting all the gaps, we have plotted in Fig. 5 the effective number of days during which Fermi/LAT was not pointed to the Sun. For the pre-anomaly period, we have taken 2017 as the representative year for reference. In the pre-anomaly years, the total Sun exposure per year is only ~90 days, counting all the Sun exposures (typically ~30 min per 96-min orbit) with effective LAT area ≥ 600 cm$^2$. The anomaly year 2018 had a larger reduction because of the planning activities for new observing modes suitable for the SODA situation (the total annual Sun exposure was ~50 days). In the years after 2018, the Sun exposure was between 60 and 65 days, which corresponds to a reduction of ~ 30% compared to pre-anomaly years.

During each of the gap intervals, we identified and counted the number of energetic events FW CMEs, DH type II bursts, and large SEP events in the appropriate catalogs available at the CDAW data center (https://cdaw.gsfc.nasa.gov). The DH type II bursts and SEP events identified in the gaps are marked with red symbols in the bottom panel of Fig. 3, while the black symbols denote events occurring outside of LAT gaps. We can see a large number of DH type II bursts and SEP events in SC 25 as was also listed in Table 1. During the entire year 2023, there was only one SGRE event (2023 December 31 event that extended to 2024 January 1), although there were many DH type II bursts and SEP events. There were 9 LAT gaps in this year, and each gap had at least one DH type II burst with a total number of 28. On the other hand, we counted only 7 large SEP events during 3 gaps; 6 gaps did not have any SEP event, although some occurred during high background from previous events. There were 55 DH type II bursts and 12 SEP events that occurred outside of the gaps in 2023. The number of LAT gaps is similar in the year 2024, the year of second SSN peak. However, there were several SGRE events in 2024 when LAT had Sun exposure. Table 2 lists the number of FW CMEs, DH type II bursts, and large SEP events that occurred during the gaps and outside the gaps. We see that most of the energetic events occurred during the solar maximum years (2022-2024). Therefore, we expect more SGRE events during these maximum years.

*Table 2. Number of LAT data gaps and energetic events in solar cycle 25*

| Year | #SGRE | #LAT DG | #FW CMEs | #IP Type II | #SEP Events |
|------|-------|---------|----------|-------------|-------------|
| 2018 | 0 | 7 | 0 (0) | 0 (0) | 0 (0) |
| 2019 | 0 | 8 | 0 (0) | 1 (0) | 0 (0) |
| 2020 | 0 | 8 | 2 (0) | 2 (0) | 0 (0) |
| 2021 | 2 | 8 | 11 (1) | 15 (4) | 2 (0) |
| 2022 | 4 | 8 | 61 (19) | 34 (8) | 5 (0) |
| 2023 | 1 | 9 | 86 (30) | 55 (28) | 12 (7) |
| 2024 | 8 | 9 | 108 (35) | 77 (39) | 16 (6) |

### 4. Estimating the number of SGRE events in SC 25

We now estimate the total number of SGRE events in SC 25 including those observed during and outside of the LAT data gaps. (i) Let us look at the 30% overall reduction in Sun exposure in SC 25. If we assume that SGRE events in SC 25 occurred with the same rate as in SC 24, we expect about 30% more events in SC 25, viz., ~33 events. (ii) The estimate in (i) does not take in to account the fact that solar activity is stronger in SC 25. From Table 1, we see that the SSN increased by 39%. One can argue that the number of SGRE events must be proportionately higher in SC 25 based on the fact that energetic CMEs generally originate from active regions. The normalized number of SGRE events SC 24 is 0.44. Multiplying this by 1.39, we get the normalized number of SGRE events in SC 25 as 0.61. This gives the total number of SGRE events in SC 25 to be ~48 (0.61 × 79.0). This indicates that ~33 events must have occurred during LAT gaps in SC 25. (iii) One of the main CME properties relevant to SGRE is that the underlying CMEs are energetic (FW). In SC 24, 17% of FW CMEs were associated with SGRE. If the association rate remains the same in SC 25, out of the 268 FW CMEs observed in SC 25, ~46 should have resulted in SGRE. This number is not too different from the estimate using SSN increase. (iv) DH type II bursts have been shown to have a quantitative relationship with SGRE events in terms of their duration and ending frequency [Gopalswamy et al. 2018]. The number of DH type II bursts in SC 25 nearly doubled (183) compared to that (99) in SC 24. If the 25% association rate of DH type II bursts with SGRE in SC 24 holds for SC 25, we expect ~46 SGRE events. This number is the same as that derived from FW CMEs because FW CMEs and DH type II bursts are closely related.

If we apply the rates in (iii) and (iv) to the number of FW CMEs and DH type II bursts in the LAT gaps, we get only ~15 and ~20 SGRE events, respectively in the gaps. Combining these numbers with the 15 observed SGRE events, we get 30 and 35 events based on the FW CME and DH type II observations. Interestingly, these numbers agree with the estimate based on the reduction in Sun exposure (i). We must point out that we have not examined other smaller (non-SODA) Fermi/LAT data gaps that are not part of the SODA-related data gaps. Therefore, the estimates from the SODA-related gaps alone are likely underestimated. There are other ways to characterize the SGRE association rate based on DH type II bursts. For example, Share et al. [2018] reported that almost all SGRE events are associated with >100 keV hard X-ray bursts during the eruption. Results of such an investigation will be reported elsewhere.

## 5. Summary and Conclusions

We investigated the solar cycle variation of the number of SGRE events during the first 61 months in SC 25 as compared to a corresponding epoch in SC 24. We attribute the primary reason for the reduction to the reduced solar coverage by Fermi/LAT following the solar array drive anomaly that started in March 2018. We estimated that the solar coverage decreased by ~30% after the anomaly. We identified 8-9 Fermi/LAT data gaps per year related to the drive anomaly. The gap intervals typically lasted for 1-3 weeks until the end of 2024 during which fermi/LAT did not observe the Sun. However, 85 FW CMEs and 79 DH type II bursts occurred during these gaps, mostly during the maximum years 2021-2024 suggesting that more SGRE events must have occurred during these gaps. The lowest estimate of the total number of SGRE events (33) was obtained based on the reduction in Sun exposure. Estimates based on the increase in SSN, FW CMEs, and DH type II bursts are 48, 46, and 46, respectively. All these estimates do indicate that SC 25 is stronger than SC 24 from the SGRE point of view as well.

## Acknowledgments

This work was supported by NASA's STEREO project, the Living with a Star program, and the Internal Science Funding Model project. HX, SA, and SY are partially supported by NSF grant AGS-2228967. We thank Elizabeth Hays, Fermi Project Scientist, for discussion on the 2018 solar array drive anomaly.